\documentclass[aps,prd,superscriptaddress,amssymb,twocolumn,eqsecnum,showpacs,epsfig,nofootinbib]{revtex4}

\usepackage{graphicx}
\usepackage{bm}
\usepackage{dcolumn}
\usepackage{amsmath}

\usepackage{amsmath}
\usepackage{amssymb}

\numberwithin{equation}{section}
\usepackage{epsfig}

\begin{document}
\newcommand{\newc}{\newcommand}

\newc{\be}{\begin{equation}}
\newc{\ee}{\end{equation}}
\newc{\bear}{\begin{eqnarray}}
\newc{\eear}{\end{eqnarray}}
\newc{\bea}{\begin{eqnarray*}}
\newc{\eea}{\end{eqnarray*}}
\newc{\D}{\partial}
\newc{\ie}{{\it i.e.} }
\newc{\eg}{{\it e.g.} }
\newc{\etc}{{\it etc.} }
{\newc{\etal}{{\it et al.}}
\newc{\lcdm}{$\Lambda$CDM}
\newcommand{\nn}{\nonumber}
\newc{\ra}{\rightarrow}
\newc{\lra}{\leftrightarrow}
\newc{\lsim}{\buildrel{<}\over{\sim}}
\newc{\gsim}{\buildrel{>}\over{\sim}}
\newcommand{\mincir}{\raise
-3.truept\hbox{\rlap{\hbox{$\sim$}}\raise4.truept\hbox{$<$}\ }}
\newcommand{\magcir}{\raise
-3.truept\hbox{\rlap{\hbox{$\sim$}}\raise4.truept\hbox{$>$}\ }}


\title{Growth index of matter perturbations in running vacuum models}

\author{Spyros Basilakos}\email{svasil@academyofathens.gr}
\affiliation{Academy of Athens, Research Center for Astronomy and
Applied Mathematics,\\
 Soranou Efesiou 4, 11527, Athens, Greece}
\author{Joan Sol\`a} \email{sola@ecm.ub.edu}
\affiliation{High Energy Physics Group, Departament d'Estructura i
Constituents de la Mat\`eria, and Institut de Ci\`encies del Cosmos (ICC),
Univ. de Barcelona, Av. Diagonal 647 E-08028 Barcelona, Catalonia, Spain}

\begin{abstract}
We derive for the first time the growth index of matter perturbations
of the FLRW flat cosmological models
in which the vacuum energy depends on redshift.
A particularly well motivated model
of this type is the so-called quantum field vacuum, in which apart from a leading constant term $\Lambda_0$ there is also an  $H^{2}$ 
dependence in the functional form of vacuum, namely $\Lambda(H)=\Lambda_{0}+3\nu (H^{2}-H^{2}_{0})$.
Since $|\nu|\ll1$ this form endows the vacuum energy of a
mild dynamics which  affects the evolution of the main cosmological
observables at the background and perturbation levels.
Specifically, at the perturbation level we find that the growth index
of the running vacuum cosmological model
is $\gamma_{\Lambda_{H}} \approx \frac{6+3\nu}{11-12\nu}$ and thus it nicely extends analytically the result
of the $\Lambda$CDM model, $\gamma_{\Lambda}\approx 6/11$.

\end{abstract}
\maketitle

\section{Introduction}
Over the last two decades, studies in cosmology strongly indicates that
we are living in a spatially flat universe
that contains $\sim 4\%$ baryonic matter, $\sim 26\%$
dark matter and $\sim 70\%$ some sort of dark energy (hereafter DE)
endowed with large negative pressure. The DE component
plays a vital role in the cosmic history because it provides
the necessary theoretical platform toward describing
the accelerated expansion of the
Universe (see Refs.\cite{Hicken2009,Ade2015} and references
therein).

Although there is a general agreement concerning the
main ingredients of the Universe, there are different proposals
regarding the underlying physics which triggers
the late cosmic acceleration. In fact the unknown nature of DE,
which challenges the foundations of theoretical physics,
has given rise to a plethora of cosmological models.
The majority of such scenarios is based either on the existence of
new fields in nature or in some modification of the theory of gravity
(for review see Ref.\cite{Amendola2010})
with the present accelerating epoch appearing as a sort of geometric effect.

An alternative avenue that one can follow
in order to explain the cosmic acceleration as well as to
overcome, or at least to alleviate, the various cosmological puzzles
is to consider a running vacuum $\Lambda(H)$. The idea was proposed some time ago\,\cite{ShapSol00,SolStef05,Babic02,Sola2007,ShapSol09} and it might be the seed of future important developments of cosmology at a more fundamental level --  see e.g. \cite{Sola2013,SolaGomez2015} for a review. This point of view is quite general and can be applied to the entire history of the Universe, including inflation and the ``graceful exit'' problem in the early Universe\,\cite{RunningInflation}. In this context,
we do not need to introduce in the analysis new fields 
in the analysis nor modify the theory of standard gravity 
[General Relativity (GR)].
In this cosmic ideology the DE equation
of state parameter
$w\equiv P_{DE}/\rho_{DE}$
is by definition identical to -1, but the vacuum energy density
is a function of time.
Notice that there is an extensive old literature on purely phenomenological models in which the cosmological term is a function of time
\cite{OzerTaha87,FreeseET87,Bertolami86,CarvalhoET92,Arc94,OverduinCooper98}, and a more recent series of works in which there is an attempt to connect the time evolution with fundamental aspects of quantum field theory (QFT)  in curved spacetime, which include the aforementioned references on running vacuum and others such as \cite{SS04,RGTypeIa,BPS09,Grande11,BasSol2014,GoSolBas2015,GomezSola2015,Basrev,SolaGomezCruz2015}.
In this extensive body of literature the time-evolving vacuum has
been phenomenologically explored as a function of time in various
possible ways, and in the more formal QFT approach it has been mainly investigated as a possible function of the Hubble
parameter. The latter is the basis for what we will refer to as the  quantum field vacuum model, in which $\Lambda=\Lambda(H)$ and on which we shall mainly concentrate here.

To test the above cosmological possibilities,
it has been proposed that
the so-called growth index, $\gamma$, of matter perturbations\,\cite{Peeb93}
can be used as an observational tool to discriminate
between modified gravity models and scalar field DE
models which obey general relativity.
Nowadays the accurate estimation of $\gamma$ is considered one of
the most basic tasks in cosmological studies.
Not surprisingly  it has become traditional to
study, for each proposed cosmological model, its background expansion as well as the growth index of matter perturbations, as in this way one may get an impression of the main cosmological and astrophysical
consequences of the model. 
For example, it has been found that
for those DE models based on GR and characterized by
a constant equation of state parameter, the asymptotic value of the
growth index is $\gamma \simeq \frac{3(w-1)}{6w-5}$
\cite{Silv94,Wang98,Linder2007,Nes08}.
Obviously, for the concordance $\Lambda$CDM model ($w=-1$) we recover
the nominal value, namely
$\gamma \approx 6/11$. As far as the modified gravity models
are concerned the situation is as follows. In the case of
the braneworld gravity of \cite{DGP} we have $\gamma \approx 11/16$
(see \cite{Linder2007,Wei08,Gong10,Fu09}),
for some $f(R)$ gravity models it has been found that
$\gamma \simeq 0.415-0.21z$
for various parameter values (see
\cite{Gann09,Tsu09}) and
finally for the Finsler-Randers cosmological model
Basilakos \& Stavrinos \cite{Basilakos2013}
have shown that $\gamma \approx 9/14$.

In this work, we wish to investigate the growth index of the
running vacuum model $\Lambda(H)=\Lambda_0+ 3\nu\,(H^{2}-H_0^2)$. The denomination of running is related to the fact that it can be motivated within the context of QFT in curved spacetime and specifically from the point of view of the renormalization group approach\, \cite{ShapSol00,SolStef05,Babic02,Sola2007,ShapSol09}, see \cite{SolaGomez2015} and\,\cite{Sola2013} for recent reviews and references therein. This running vacuum model generalizes the traditional $\Lambda$CDM model at the background level and can be put to the test. Currently the value of the dimensionless free parameter $\nu$  is observationally allowed to be in the ballpark of $|\nu|\sim {\cal O}(10^{-3})$\,\cite{SS04,RGTypeIa,BPS09,Grande11,GoSolBas2015,GomezSola2015,Basrev,SolaGomezCruz2015}. From its theoretical interpretation [namely, as being the coefficient of the $\beta$-function of the running $\Lambda(H)$] it is a natural value, which in addition fits in with the existing theoretical estimates\,\cite{Sola2007,Sola2013}.

To the best of our knowledge, we are unaware of any previous
analysis of this kind applied to dynamical vacuum models and for this reason we consider that it can be of theoretical interest, and maybe we can extract also some practical consequences.

The structure of the article is as follows.
Initially in section 2, we briefly present the background
cosmological equations. The basic theoretical elements of
the linear growth are discussed in section 3, while in section
4 we provide the growth index analysis in the case of
the running vacuum. In section 4 we compare different $\gamma(z)$
parametrizations and, finally, in section 5 we provide some discussion and finish with our main conclusions.

\section{Background evolution}
The physics of the dynamical vacuum model under consideration
is based on the renormalization group (RG) in QFT according to the aforementioned references.
Within this framework, the evolution of the
vacuum in the current epoch is given by
\be
\label{RGlaw2}
\Lambda(H)=\Lambda_0+ 3\nu\,(H^{2}-H_0^2)\,,
\ee
where $\Lambda_0\equiv\Lambda(H_0)=3\Omega_{\Lambda 0}H^{2}_{0}$ and
$\nu$ is provided in the RG
context as a ``$\beta$-function which determines the running of the cosmological ``constant'' (CC)  within QFT in curved spacetime\,\cite{Sola2013}.
The value of $\nu$ is estimated through the upper bound
$|\nu|\lesssim 1/(12\pi)\simeq 2.6\times 10^{-2}$, which is approximate and is valid as an order of magnitude. It ensues from assuming that the effective masses of the heaviest particles involved in the loops for calculating the $\beta$-function is of the order of the Planck mass\,\cite{ShapSol00,SS04,RGTypeIa}, but in general it is not completely fixed. If there is a large multiplicity of heavy particles at a grand unified scale below the Planck mass, it could as well stay of order $10^{-2}$. Notice that the observational bound on $\nu$ depends on the particular implementation of the model (\ref{RGlaw2}), namely, on whether e.g., the vacuum exchanges energy with matter or not. In the simplest cases the natural theoretical estimate yields
$\nu=10^{-5}-10^{-3}$ (see Sol\`a's Ref.\,\cite{Sola2007} for details) and in these cases $\nu$ is generally significantly smaller than the original upper bound. At the same time
using cosmological data it has been found that
$|\nu|={\cal O}(10^{-3})$ \cite{BPS09,Grande11,GoSolBas2015,GomezSola2015,Basrev,SolaGomezCruz2015}, which
is  in agreement with the aforementioned theoretical expectations. Let us, however, point out that in models in which matter and DE are self-conserved the observational limits are weaker and they tolerate the order of magnitude $\nu\sim 10^{-2}$ from the original estimate, as shown in the recent work \cite{GKS2015}.

Dynamically speaking, since $|\nu|\ll 1$ in all theoretical implementations, it is easy to check that
prior to the present epoch
the low-energy behavior of the model
tends to the usual $\Lambda$CDM model, but it
is by no means identical (for a recent review see
Refs.~\cite{SolaGomez2015,Basrev} and references therein).

Considering  Eq.(\ref{RGlaw2}), let us now
focus on the derivation of the Friedmann equations. Such a proceduce
is perfectly allowed by the cosmological principle
embedded in the FLRW metric. Namely, the $\Lambda$ term may perfectly evolve with the cosmic expansion, meaning that ultimately evolves with the cosmic time, $t$, but in general it depends on an intermediate variable, which in our case is the Hubble function, $H=\dot{a}/{a}$, where $a(t)$ is the scale factor and the overdot denotes a derivative with respect to $t$. The corresponding 
generalization of the Friedmann equations reads:
\begin{eqnarray}
8\pi G \rho_{\rm tot}&\equiv& 8\pi G \rho_{m}+\Lambda(H)
=3H^{2}\;,
\label{friedr}\\
8\pi G P_{\rm tot}&\equiv &8\pi G P_{m}-\Lambda(H) =-2{\dot H}-3H^2\,,
\label{friedr2}
\end{eqnarray}
where the total energy density is $\rho_{\rm tot}=\rho_{m}+ \rho_{\Lambda}$
(with $\rho_{\Lambda}=\Lambda/8\pi G$ the vacuum component of it)
and $P_{\rm tot}=P_{m}+P_{\Lambda}$ is the total pressure. For the matter-dominated epoch, and of course also in our days,
$(P_{m},P_{\Lambda})=(0,-\rho_{\Lambda})$, where the dynamical character of the vacuum does not alter the usual equation of state that it satisfies. This is an important point to remark.
In fact, this observation explains why the dynamics of the vacuum entails a corresponding modification of the local energy conservation for matter at fixed $G$. The outcome is that matter must exchange energy with the vacuum in order to fulfill the Bianchi identity, and this translates into the following generalized conservation law involving both matter and vacuum energy densities:
\begin{equation}
\dot{\rho}_{m}+3H\rho_{m}=-\dot{\rho_{\Lambda}}\,. \label{frie33}
\end{equation}
This equation is actually not independent of (\ref{friedr}) and (\ref{frie34}), 
and therefore, using any two of them,
it is easy to derive the equation of motion for the Hubble rate:
\begin{equation}
\dot{H}+\frac{3}{2}H^{2}=4\pi G \rho_{\Lambda}=
\frac{\Lambda(H)}{2} \;.
\label{frie34}
\end{equation}
From (\ref{frie34}) and the vacuum model equation (\ref{RGlaw2}), we are able to determine the explicit form of $H$ as a function of time:
\begin{equation}
\label{frie455} H(t)=H_{0}\,\sqrt{\frac{\Omega_{\Lambda 0}-\nu}{1-\nu}} \;
\coth\left[\frac32\,H_{0}\sqrt{(\Omega_{\Lambda 0}-\nu)(1-\nu)}\;t\right]\,,
\end{equation}
where $\Omega_{\Lambda 0}=1-\Omega_{m0}$ and
$H_{0}$ is the Hubble constant.
Utilizing $H={\dot a}/a$ the cosmic time, $t(a)$, follows:
\begin{equation}
t(a)=\frac{2}{3\,H_{0}}\, \frac{{\rm sinh^{-1}} \left(\sqrt{ \frac{\Omega_{\Lambda 0}-\nu}
{\Omega_{m 0}}} \;a^{3(1-\nu)/2} \right)}{\sqrt{(\Omega_{\Lambda 0}-\nu)(1-\nu)}}\,.
\label{frie456t}
\end{equation}
{Inverting Eq.(\ref{frie456t}) we easily determine the scale factor $a=a(t)$.
Therefore, inserting Eq.(\ref{frie456t}) into
Eq.(\ref{frie455}) we arrive at
\begin{equation}
\label{anorm11}
E^{2}(a)\equiv\frac{H^{2}(a)}{H^{2}_{0}}=
{\tilde \Omega}_{\Lambda 0}+{\tilde \Omega_{m0}}a^{-3(1-\nu)}\;,
\end{equation}
where we have rescaled
\begin{equation}
\label{otran1}
{\tilde \Omega_{m 0}}\equiv \frac{\Omega_{m0}}{1-\nu}\,,\ \ \
{\tilde \Omega_{\Lambda 0}}\equiv \frac{\Omega_{\Lambda 0}-\nu}{1-\nu}\,.
\end{equation}
Notice that for $\nu=0$ all the above formulas correctly reduce to the standard ones for the $\Lambda$CDM, and the rescaled parameters become the ordinary ones, $\tilde{\Omega}_{i0}\to\Omega_{i0}$. Moreover,
whether in rescaled form or not the  cosmological parameters obey  the
standard cosmic sum rule, namely
$\tilde{\Omega}_{m0}+\tilde{\Omega}_{\Lambda 0}=1=\Omega_{m 0}+\Omega_{\Lambda 0}$.

Regarding the matter evolution, from Eqs. (\ref{friedr}) and (\ref{friedr2}) we find $\dot{H}=-4\pi\,G\rho_m$, and  combining the latter with Eqs.(\ref{RGlaw2}) and (\ref{frie33}) we obtain a differential equation for the matter density:
$\dot{\rho}_m+3H\rho_m=3\nu H\rho_{m}$. Integrating it
(using $\dot{\rho}_m=aH d\rho_m/da$) we find
\begin{equation}\label{mRG}
\rho_m(a) =\rho_{m0}\,a^{-3(1-\nu)}\,.
\end{equation}
Notice, that $\rho_{m0}$ is the matter density at the present time ($a=1$),
and therefore $\Omega_{m0}=\rho_{m0}/\rho_{c0}$, where
$\rho_{c0}=3H_0^2/8\pi G$ is the current critical density. 
As expected, we recover the standard matter 
conservation law $\rho_m\sim a^{-3}$ only for $\nu=0$. 
However, thanks to $\nu\neq 0$ we can have a mild 
dynamical vacuum evolution; see Eq.\,(\ref{RGlaw2}).

Defining $\Omega_m(a)\equiv{\rho_m(a)}/{\rho_c(a)}$ it is
easy to obtain, with the aid of Eqs.(\ref{mRG}) and
(\ref{anorm11}),
\begin{equation}\label{aeffeom}
\Omega_m(a)=\frac{\Omega_{m0}a^{-3(1-\nu)}}{E^2(a)}\,.
\end{equation}

For convenience we also define
\begin{equation}\label{effeom}
{\tilde
\Omega}_m(a)=\frac{\tilde{\Omega}_{m0}a^{-3(1-\nu)}}{E^2(a)}=\frac{\Omega_{m}(a)}{1-\nu}\,.
\end{equation}
Differentiating Eq.(\ref{effeom}) and utilizing
(\ref{anorm11}) we find that
\be
\label{domm}
\frac{d{\tilde \Omega}_{m}}{d{\ln }a}=
-3(1-\nu){\tilde \Omega}_{m}(a)\left[1-{\tilde \Omega}_{m}(a)\right]\;.
\ee
Subsequently, upon substituting Eq.(\ref{mRG}) into Eq.(\ref{frie33})
and integrating
once more in the scale factor variable, we are led to
the evolution of the vacuum energy density:
\begin{equation}
\label{CRG}
\rho_{\Lambda}(a)=\rho_{\Lambda 0}+\frac{\nu\,\rho_{m0}}{1-\nu}\,
\left[a^{-3(1-\nu)}-1\right]\,.
\end{equation}
Once more for $\nu=0$ the cosmological
solutions of the running vacuum model under study
boil down to the concordance
$\Lambda$CDM cosmology, and in this case $\rho_{\Lambda}=\rho_{\Lambda 0}$ at all times.

Finally, the observational viability of the current vacuum
model has been tested previously, and in the most recent analysis
provided in G\'omez-Valent et al. \cite{GoSolBas2015}
it is found that
$(\Omega_{m},\nu)=(0.282\pm0.012,0.0048\pm 0.0032)$
(see also Table 1 in Ref.\cite{Basrev}). For the rest of the
paper we shall take this result as the basis for our estimates.
Notice that, due to the rescaling
[see the first equality of (\ref{otran1})] we have
${\tilde \Omega}_{m 0}=0.283\pm 0.012$.}
Recall that
for the concordance $\Lambda$ model ($\nu=0$)
$\Omega_{m0}^{(\Lambda)}=0.291 \pm 0.011$ which
is in agreement with the recent {\it Planck 2015} results \cite{Ade2015}.

\section{Linear Growth}
In this section we concentrate on
the basic linear equation that governs the evolution of the matter
perturbations. Following
\cite{Arc94,BPS09,GoSolBas2015}
we write the following equation for the matter density contrast
$D\equiv\delta\rho_m/\rho_m$:
\begin{equation}
\label{eq:11}
\ddot{D}+(2H+Q)\dot{D}-\left(4\pi G \rho_{m}
-2HQ-\dot{Q} \right)D=0,
\end{equation}
where $Q(t)=-\dot{\rho}_{\Lambda}/\rho_{m}$. For a formal proof of this equation in relativistic cosmology, see Refs.\cite{BasSol2014,GoSolBas2015}. It assumes that the DE perturbations are very small and that the divergence of the perturbed matter velocity is also negligible. Obviously, the running vacuum
energy still affects the growth factor through the function
$Q(t)$, and therefore, it affects the background evolution of the matter perturbations.

In this context, we can rewrite the homogeneous form of
Eq.(\ref{eq:11}) in terms of the scale factor (using $d/dt=aH(a) d/da$)
as follows:
\bear
\label{diffeqDa}
&&a^{2}\frac{d^{2}D}{da^{2}}+\left(3a+a\frac{d{\ln}H}{d{\rm ln}a}+\frac{aQ}{H}\right)\,\frac{dD}{da}\; \nonumber \\
&&=\left(\frac{3}{2}\Omega_{m}-\frac{2Q}{H}-\frac{a}{H}\frac{dQ}{da}\right)D\; \nonumber \\
&&=\left[\frac{3}{2}(1-\nu){\tilde \Omega}_{m}-\frac{2Q}{H}-\frac{a}{H}\frac{dQ}{da}\right]D\,,
\eear
where
\be
\label{EE1}
\frac{d{\ln}H}{d{\rm ln}a}=\frac{d{\ln}E}{d{\rm ln}a}=
-\frac{3}{2}(1-\nu){\tilde \Omega}_{m}(a)\;,
\ee
\be
\label{EE2}
\frac{Q(a)}{H(a)}=-\frac{ {\dot \rho}_{\Lambda}(a) }{H(a)\rho_{m}(a)}=3\nu
\ee
and
\be
\label{EE3}
\frac{a}{H(a)}\frac{dQ}{da}=-\frac{9}{2}\nu(1-\nu){\tilde \Omega}_{m}(a) .
\ee
Notice, that in order to derive the above expressions
we have utilized Eqs.(\ref{anorm11})-(\ref{CRG}).
The growing mode solution of Eq.(\ref{diffeqDa}) is written as
(for more details see Ref. \cite{BPS09})
\begin{equation}
\label{solchangextoa}
D(a)=C_1 a^{\frac{9\xi-4}{2}}E(a)\;
F\left(\frac{1}{3\xi}+\frac{1}{2},\, \frac{3}{2},
\,\frac{1}{3\xi}+\frac{3}{2};\,-\frac{{\tilde \Omega}_{\Lambda 0}}{{\tilde \Omega}_{m0}}\,\,a^{3\xi}
\right)\,,
\end{equation}
where $\xi=1-\nu$, $C_1$ is an integration
constant to be constrained by an initial condition,
and $F$ is the hypergeometric function \,\,\cite{Gradshteyn}.

Of course for the concordance $\Lambda$CDM model [$\nu=0$ or $Q(t)=0$]
Eq.(\ref{diffeqDa}) reduces to the
standard perturbation equation a solution
of which is (see Refs. \cite{Peeb93,Linjen03})
\be\label{eq24}
D_{\Lambda}(a)=\frac{5\Omega_{m0}
  E(a)}{2}\int^{a}_{0}
\frac{dx}{x^{3}E^{3}(x)} \;\;.
\ee

\section{Growth index}
For any type of dark energy, a useful parametrization
of the matter perturbations is based on the growth rate of clustering\,\cite{Peeb93}. In our framework, the natural parametrization is\footnote{In the running vacuum model (\ref{RGlaw2})
one may check from (\ref{aeffeom}) that, at large redshifts $z\gg 1$,
$\Omega_{m}(z) \sim 1-\nu$. Therefore, if the growth rate of clustering is modeled as a power law, then
it is more appropriate to use $f(a)\simeq {\tilde \Omega}_{m}(z)^{\gamma}$, because for $z\gg 1$ we achieve the correct normalization ${\tilde \Omega}_{m}(z) \sim 1$.}
\be
\label{fzz221}
f(a)=\frac{d{\rm ln} D}{d{\rm ln} a}\simeq {\tilde \Omega}^{\gamma}_{m}(a) \;,
\ee
with $\tilde{\Omega}_{m}(a)$ defined in Eq.\,(\ref{effeom}).
The exponent $\gamma$ is the so-called growth index
(see Refs.~\cite{Silv94,Wang98,Linjen03,Linder2007,Nes08,Lue04})
and it plays an important role in cosmological studies as we discussed in the
Introduction.
Inserting the first equality of (\ref{fzz221})
into Eq.(\ref{diffeqDa})
we derive after some calculations 
$$\frac{df}{d{\rm ln}a}+
\left(2+\frac{Q}{H}+\frac{d{\ln}H}{d{\rm ln}a}\right)f+f^{2}
$$
\be\label{fzz222}
=\frac{3(1-\nu){\tilde \Omega}_{m}}{2}-\frac{2Q}{H}-\frac{a}{H}\frac{dQ}{da} \;.
\ee
or
\bear
\label{fzz223}
&&\frac{df}{d{\rm ln}a}+
\left[\frac{1}{2}+\frac{3}{2}(1-\nu){\tilde \Omega}_{\Lambda}+\frac{9\nu}{2}\right]f+
f^{2}\; \nonumber \\
&&=\frac{3(1-\nu){\tilde \Omega}_{m}}{2}-6\nu+\frac{9\nu(1-\nu)}{2}{\tilde \Omega}_{m}
\eear
where ${\tilde \Omega}_{\Lambda}(a)\equiv1-{\tilde \Omega}_{m}(a)$.
At this point it is interesting to mention that there
have been many theoretical
speculations concerning the functional form of the growth index
and indeed various candidates have been proposed in the literature.
In this work, we phenomenologically parametrize $\gamma(a)$
by the following general relation \cite{Pour}
\be
\label{gzzz}
\gamma(a)=\gamma_{0}+\gamma_{1}y(a)\;.
\ee
In other words, Eq.(\ref{gzzz}) can be seen
as a first-order Taylor expansion
around some cosmological function $y(a)$. The following options have been considered in the literature:
$\omega(a)={\rm ln}{\tilde \Omega}_{m}(a)$ 
(hereafter $\Gamma_{0}$ parametrization: \cite{Steig}),
$a(z)$ (hereafter $\Gamma_{1}$ parametrization: \cite{Bal08})
and $z$ (hereafter $\Gamma_{2}$ parametrization: \cite{Pol}).
Below we briefly present
various forms of $\gamma(a)$ for the running vacuum model in these various parametrizations.

\subsection{$\Gamma_{0}$ parametrization}
In this parametrization we use $y(a)=\omega={\rm ln}{\tilde \Omega}_{m}(a)$.
Obviously, for $z\gg 1$, namely ${\tilde \Omega}_{m}(a)\to 1$ (or $\omega \to 0$)
the asymptotic value of the growth index becomes
$\gamma_{\infty}\approx \gamma_{0}$.
Steigerwald et al. \cite{Steig} proposed a general mathematical treatment
[see their Eqs. (5)-(12)]\footnote{The
methodology of Steigerwald et al. \cite{Steig} can be applied to the framework
of $\gamma(a)=\sum_{n=0}^{N} \gamma_{n}\frac{\omega^{n}(a)}{n!}$. However, for the
purpose of our study we restrict our analysis to $N=1$.
We would like to point
that Eqs.(3) and (5) of Steigerwald et al. \cite{Steig}
have a typo. Indeed one has
to replace there the quantity $1+\nu_{H}$ with $2\nu_{H}$. Notice, however
that this typo does not alter our results because in our case the coefficient
$\nu_{H}$ is a constant [see Eq.(\ref{nn1})] which implies 
that $N_{n}=0$, see Eq.(\ref{Coef}).}
which provides compact
analytic formulas for the
coefficients $\gamma_{0}$ and $\gamma_{1}$.
{ These authors start from the
fact that for a large family of dark energy models (including those of
modified gravity)
the linear differential equation of the matter
perturbations takes on the form
\begin{equation}
\label{eq:111}
\ddot{D}+2\nu_{H}H\dot{D}-4\pi G\mu \rho_{m} D=0 \;.
\end{equation}
Naturally, any modification to the Friedmann equation and to the
theory of gravity is included in the quantities $\nu_{H}$ and $\mu$.
For the nominal scalar field dark energy which adheres to general relativity
we have $\nu_{H}=\mu=1$, while for modified gravity models
we get $\nu_{H}=1$ and $\mu \ne 1$. Notice that for the latter cases
we have ${\tilde \Omega}_{m}(a)\equiv \Omega_{m}(a)$ by definition. 
In the following  we will determine the precise relation between the generic coefficients $(\nu_H,\mu)$ and our vacuum parameter $\nu$.

To start with, we expect that if we allow interactions in the dark
sector, then, in general, the cosmological quantities
$(\nu_{H},\mu)$ are different from unity. Furthermore, if the matter
component evolves differently from the usual power law $a^{-3}$ one
may expect that the quantity $\Omega_{m}(a)=\rho_{m}(a)/\rho_{c}(a)$
is slightly different from ${\tilde \Omega}_{m}(a)$  as defined in
Eq.\,(\ref{effeom}). For example, in our case, due to
Eqs.(\ref{effeom}) and the first equality of (\ref{fzz221}) one can
write Eq.(\ref{eq:111}) as follows:\be \label{fzz444} \frac{df}{d{\rm ln}a}+
\left(2\nu_{H}+\frac{d{\ln}H}{d{\rm ln}a}\right)f+f^{2}
=\frac{3{\tilde \mu} {\tilde \Omega}_{m}}{2}\;, \ee where ${\tilde
\mu}\equiv \mu (1-\nu)$.} Similar to \cite{Steig} let us transform
Eq.(\ref{fzz444}) as \be
\frac{d\omega}{d{\ln}a}(\gamma+\omega\frac{d\gamma}{d\omega}) +{\rm
e}^{\omega \gamma}+2\nu_{H}+\frac{d{\ln}H}{d{\rm
ln}a}-\frac{3}{2}{\tilde \mu} {\rm e}^{\omega(1-\gamma)}=0 \ee where
we have set $\omega={\rm ln}{\tilde \Omega}_{m}(a)$. Within the
mathematical framework of \cite{Steig} one finds \be \label{g000}
\gamma_{0}=\frac{3(M_{0}+M_{1})-2({\cal
H}_{1}+N_{1})}{2+2X_{1}+3M_{0}} \ee and \bear \label{g111}
&&\gamma_{1}=3\frac{M_{2}+2M_{1}B_{1}(1-y_{1})+M_{0}B_{2}(1-y_{1},-y_{2})}
{2(2+4X_{1}+3M_{0})}\; \nonumber \\
&&-2\frac{B_{2}(y_{1},y_{2})+X_{2}\gamma_{0}+{\cal H}_{2}+N_{2}}
{2(2+4X_{1}+3M_{0})}\,.
\eear
The following quantities have been defined:
\be
X_{n}=\left.\frac{d^{n}(d\omega/d{\rm ln}a)}{d\omega^{n}}\right|_{\omega=0}\,,
\ \
M_{n}=\left.\frac{d^{n}{\tilde \mu}}{d\omega^{n}}\right|_{\omega=0}
\ee
and
\be \label{Coef}
N_{n}=\left.\frac{d^{n}\nu_{H}}{d\omega^{n}}\right|_{\omega=0}\,,\ \
{\cal H}_{n}=\left.\frac{d^{n}(d{\rm ln}H/d{\rm ln}a)}{d\omega^{n}}\right|_{\omega=0}\,,
\ee
with  $\frac{d^{0}}{d\omega^{0}}\equiv 1$.
Also $B_{1,2}$ are the Bell polynomials of first and second kind, namely
$B_{1}(y_{1})=y_{1}$ and $B_{2}(y_{1},y_{2})=y_{1}^{2}+y_{2}$. According to
Steigerwald et al. \cite{Steig} [see their Eq.(10)] the pair
$(y_{1},y_{2})$ is equal to $(\gamma_{0},0)$ implying
$B_{1}(1-y_{1})=1-\gamma_{0}$, $B_{2}(y_{1},y_{2})=\gamma_{0}^{2}$
and $B_{2}(1-y_{1},-y_{2})=(\gamma_{0}-1)^{2}$.

\begin{figure}
\mbox{\epsfxsize=8.2cm \epsffile{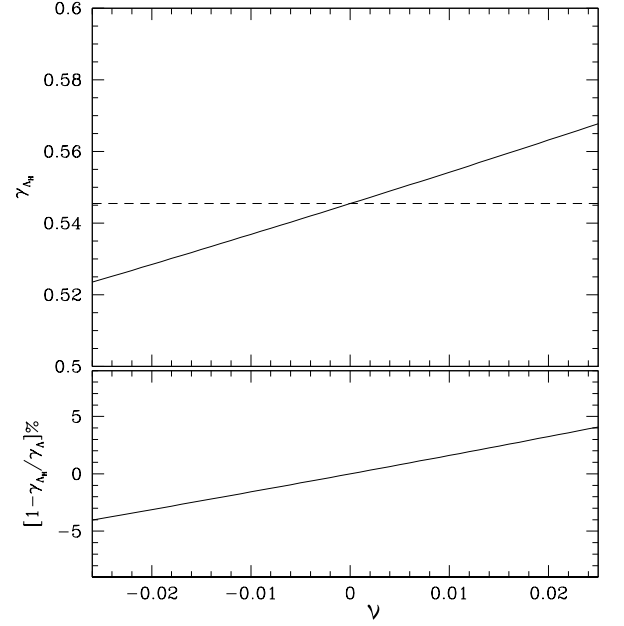}} \caption{In the
upper panel we present the asymptotic value of the growth
index as a function of $\nu$ (solid line).
Notice, that the dashed curve corresponds to $\gamma_{\Lambda}\approx 6/11$.
In the lower panel we show the relative deviation
$[1-\gamma_{\Lambda_{H}}/\gamma_{\Lambda}]\%$.}
\end{figure}

For the current running vacuum model, the quantities $\nu_{H}$ and $\tilde{\mu}$ can be easily identified  on comparing Eq.(\ref{fzz222}) with Eq.(\ref{fzz444}), and we find
\be
\label{nn1}
\nu_H=1+\frac{Q}{2H}=1+\frac32\,\nu
\ee
and
\be
\label{mm1}
{\tilde \mu}=1-\nu-\frac{4Q}{3{\tilde \Omega}_{m}H}-\frac{2a}{3{\tilde \Omega}_{m}H}\frac{dQ}{da}=
1-\nu-\frac{4\nu}{{\tilde \Omega}_{m}}+3\nu(1-\nu),
\ee
where in the derivation of the second equalities we have used
Eqs.(\ref{EE2}) and (\ref{EE3}).
Then, based on Eqs.(\ref{domm}), (\ref{EE1}), (\ref{nn1}) and
(\ref{mm1}), we obtain
$$
\{ M_{0},M_{1},M_{2},N_{1},N_{2}\}=\{ 1-2\nu-3\nu^{2},4\nu,-4\nu,0,0\}
$$
$$
\{ X_{1},{\cal H}_{1}\}=\{ X_{2},{\cal H}_{2}\}=
\{ 3(1-\nu),-\frac{3(1-\nu)}{2}\}\,,
$$
and thus
\be
\label{g002}
\gamma_{0}=\frac{6+3\nu(1-3\nu)}{11-3\nu(4+3\nu)}
\ee
\bear
\label{g112}
&&\gamma_{1}=3\frac{-4\nu+8\nu(1-\gamma_{0})+(1-2\nu-3\nu^{2})(\gamma_{0}-1)^{2}}
{2[2+12(1-\nu)+3(1-2\nu-3\nu^{2})]}\; \nonumber \\
&&-\frac{2\gamma_{0}^{2}+6(1-\nu)\gamma_{0}-3(1-\nu)}
{2[2+12(1-\nu)+3(1-2\nu-3\nu^{2})]}.
\eear
If we take the aforementioned fitting value $\nu=0.0048$ from \cite{GoSolBas2015}, we find
$(\gamma_{0},\gamma_{1})=(0.5496, -0.009)$.
For $\nu=0$ we recover the $\Lambda$CDM pair
$(\gamma_{0},\gamma_{1})=(\frac{6}{11}, -0.00729)$
as it should \cite{Steig}. Lastly, since $\nu$ is of
order of ${\cal O}(10^{-3})$
\cite{BPS09,Grande11,GoSolBas2015} it is safe to neglect high-order terms of
$\nu$ from Eq.(\ref{g002}). In that case the asymptotic
value of the growth index becomes
\be
\label{g003}
\gamma_{\infty}\approx \gamma_{0}=
\frac{6+3\nu}{11-12\nu}\simeq \frac{6}{11}\left(1+\frac{35}{22}\,\nu\right)
\ee
where $\gamma_{\infty} \approx [\gamma(a)]_{{\tilde \Omega}_{m}=1}$.

In the upper panel of Fig.1 we show the asymptotic value of the
growth index as a function of $\nu$ (solid curve), where $\nu$ parameter
lies in the
theoretical interval $[-\frac{1}{12\pi},\frac{1}{12\pi}]$ which
implies that $\gamma_{\Lambda_{H}} \in [0.5235,0.5678]$.
In the lower panel of Fig.1 we present the relative deviation
$[1-\gamma_{\Lambda_{H}}/\gamma_{\Lambda}]\%$ of the growth index with respect to
$\gamma_{\Lambda} \approx 6/11$.
We observe that for negative values of $\nu$ the asymptotic value of the growth index becomes less than $6/11$ (the opposite holds for positive values).
This deviation can reach $\sim \pm 5\%$ when we attain the aforementioned theoretical upper bound of $\nu=\pm 1/12\pi\simeq \pm0.026$. These features can be easily understood from the approximate formula on 
the rhs of Eq.\,(\ref{g003}). As advanced in Sec. 2, there 
are DE models based also on 
Eq.\,(\ref{RGlaw2}) and generalizations there (cf. Ref.\cite{GKS2015}) in 
which the observational limits lie at the 
border of the upper theoretical bound, so in general 
we can say that $\sim5\%$ corrections to $\gamma$ are 
conceivable and could perhaps be within reach in the future.  
For the cases in which Eq.\,(\ref{RGlaw2}) represents a 
dynamical vacuum model in interaction 
with matter, however, the corrections are smaller. 
Using the latest observed value of $\nu=0.0048 \pm 0.0032$ provided by
G\'omez-Valent et al. \cite{GoSolBas2015} (see also Table 1 in Ref.\cite{Basrev})
we find $\gamma_{\Lambda_{H}}=0.5496 \pm 0.0028$, to be compared with $\gamma_{\Lambda}\simeq 0.5454$ for the concordance model, hence a $\lesssim1\%$ correction.

\subsection{$\Gamma_{1,2}$ parametrizations}
Here we generalize the original Polarski \& Gannouji \cite{Pol}
work. In particular, changing the variables in Eq.(\ref{fzz223})
from $a(z)$ to redshift [$\frac{df}{da}=-(1+z)^{-2}\frac{df}{dz}$]
and utilizing $f(z)={\tilde \Omega}_{m}(z)^{\gamma(z)}$ we find
\bear
\label{Poll}
&&-{\tilde \Omega}^{\gamma}_{m}\left[
(1+z)\gamma^{\prime}{\rm ln}({\tilde \Omega}_{m})+3\gamma (1-\nu){\tilde \Omega}_{\Lambda} \right]\; \nonumber \\
&&+\left[\frac{1}{2}+\frac{3}{2}(1-\nu){\tilde \Omega}_{\Lambda}+\frac{9\nu}{2}\right]{\tilde \Omega}_{m}^{\gamma}+
{\tilde \Omega}^{2\gamma}_{m} \; \nonumber \\
&&=
\frac{3(1-\nu){\tilde \Omega}_{m}}{2}-6\nu+\frac{9\nu(1-\nu)}{2}{\tilde \Omega}_{m}
\eear
where prime denotes a derivative with respect to the redshift.
For those $y(z)$ functions which satisfy the restriction $y(z=0)=0$
[or $\gamma(z=0)=\gamma_{0}$]\footnote{For $\Gamma_{1,2}$ parametrizations
$\gamma_{0}=\gamma(z=0)$
is different with that of Steigerwald et al. \cite{Steig}, namely
$\gamma_{0}=[\gamma(a)]_{{\tilde \Omega}_{m}=1}\approx \gamma_{\infty}$ (see section 3A).},
we obtain the parameter $\gamma_{1}$ in terms of $\gamma_{0}$, $\Omega_{m0}$ and $\nu$. Specifically,
if we substitute $z=0$ and $\gamma^{\prime}(0)=\gamma_{1}y^{\prime}(0)$
in Eq.(\ref{Poll}), then we have
\be
\label{Poll2}
\gamma_{1}=\frac{{\tilde \Omega}_{m0}^{\gamma_{0}}-3(1-\nu)(\gamma_{0}-\frac{1}{2})
{\tilde \Omega}_{\Lambda 0}+\frac{1}{2}-\frac{3}{2}(1-\nu){\tilde \Omega}_{m0}^{1-\gamma_{0}}+\Psi_{0}}
{y^{\prime}(0)\ln  {\tilde \Omega}_{m0}}
\ee
where
$$
\Psi_{0}=\frac{9\nu}{2}-\frac{9\nu(1-\nu){\tilde \Omega}^{1-\gamma_{0}}_{m0}}{2}+6\nu {\tilde \Omega}_{m0}^{1-\gamma_{0}} .
$$
Clearly, in the case of the usual $\Lambda$ cosmology ($\nu=0$, $\Psi_{0}=0$)
the above formula boils down to that of
Polarski \& Gannouji \cite{Pol} for $y(z)=z$ ($\Gamma_{2}$ parametrization).
Furthermore, based on the $\Lambda$CDM cosmological model and
for $y(z)=1-a(z)=\frac{z}{1+z}$ ($\Gamma_{1}$ parametrization),
we also confirm the literature results
(see Refs. \cite{Ishak09},\cite{Bel12} and \cite{DP11}).
Now, due to the fact that
the function $y(z)=z$ goes to infinity at large redshifts
for the rest of the paper we focus on the $\Gamma_{1}$ parametrization
which implies $y^{\prime}(0)=1$. In this case
one can easily see that
$\gamma_{\infty}\simeq \gamma_{0}+\gamma_{1}$ as long as
$z\gg 1$. Therefore, plugging $\gamma_{0}=\gamma_{\infty}-\gamma_{1}$ into
Eq.(\ref{Poll2}) and using at the same time
$\gamma_{\infty}\approx \frac{6+3\nu}{11-12\nu}$ we can derive the constants
$\gamma_{0,1}$ in terms of $\Omega_{m0}$ and $\nu$. For example,
in the case of $(\Omega_{m0},\nu)=(0.283,0.0048)$, we find
$(\gamma_{0},\gamma_{1})=(0.5636,-0.140)$, while for the concordance
$\Lambda$ cosmological model we get $(\gamma_{0},\gamma_{1})=(0.5565,-0.110)$
for $(\Omega_{m0},\nu)=(0.291,0)$.

\begin{figure}[t]
\includegraphics[width=0.5\textwidth]{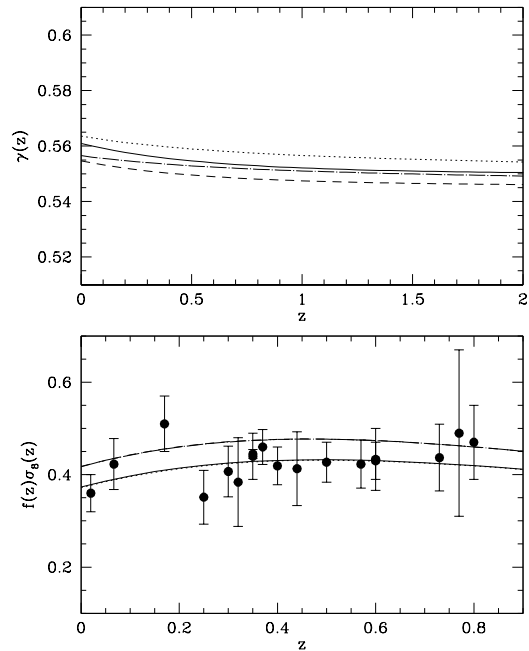}
\caption{{\em Upper Panel:} The evolution of the growth index of matter
perturbations. The curves are as follows. The results for
$\Lambda(H)$ in the $\Gamma_{0,1}$ parametrization are given by
the solid and dotted lines.
Moreover, the dashed and the dot-dashed curves
correspond to the $\Lambda$CDM in the $\Gamma_{0,1}$ parametrization
respectively. {\em Lower Panel:} Comparison of the observed and
theoretical evolution of the growth
rate $f(z)\sigma_{8}(z)$. The growth data can be found in
\cite{Ness2015} (see their Table 1). For the
$\Lambda$CDM we use $\sigma_{8}=0.829$ while for the running vacuum model
we have $\sigma_{8}=0.758$ (see \cite{GoSolBas2015,Basrev}).}
\end{figure}


\section{Discussion and Conclusions}
In the upper panel of Fig. 2
we present the evolution of the growth index
for the running vacuum model $\Lambda(H)$ in the $\Gamma_{0}$ parametrization (solid line),
and in the $\Gamma_{1}$ parametrization (dotted line).
In the same figure the dashed and the dot-dashed curves
correspond to the $\Lambda$CDM in the $\Gamma_{0}$ and $\Gamma_{1}$ parametrizations, respectively.
The comparison indicates
that the growth index of the $\Lambda(H)$ and $\Lambda$CDM
cosmological models is well approximated by the $\Gamma_{0}$ and
$\Gamma_{1}$ parametrizations. Specifically, we find that the corresponding
relative deviations are
$[1- \gamma^{(\Gamma_{1})}_{\Lambda_{H}}/ \gamma^{(\Gamma_{0})}_{\Lambda_{H}}]
\sim 0.7\%$ and
$[1- \gamma^{(\Gamma_{1})}_{\Lambda}/ \gamma^{(\Gamma_{0})}_{\Lambda}]
\sim 0.6\%$ which means that essentially both parametrizations are equivalent,
namely they provide the same growth index results.
Based on the previous analysis, it is interesting to mention
that the differences that we find
with respect to the $\Lambda$CDM growth index are near the
edge of the present experimental limits. Indeed,
in a recent analysis of the clustering properties
of luminous red galaxies and the growth rate
data provided by the various galaxy surveys, it has been
found that $\gamma=0.56\pm 0.05$ and $\Omega_{m0}=0.29\pm 0.01$
\,\cite{Athina2014}. Obviously, the prediction of $\gamma$ for all our
possible cases lies within $1\sigma$ of that range.

In the lower panel of Fig.2 we show
the growth data (solid points; see Ref.\cite{Ness2015})
together with the predicted
$f(z)\sigma_{8}(z)\simeq \sigma_{8}D(z)\Omega_{m}(z)^{\gamma(z)}$
for the running vacuum $\Lambda(H)$ in the $\Gamma_{0}$ (solid line) and
$\Gamma_{1}$ (dotted line) parametrizations, and for the
$\Lambda$CDM (dashed line, $\Gamma_{0}$ 
and dot-dashed line, $\Gamma_{1}$)
We observe that the running vacuum model reproduces the growth data well,
in a way that is compatible with the $\Lambda$CDM model.
Notice that, in order to obtain the above results 
for the concordance $\Lambda$ cosmology we utilize
$\sigma_{8}=0.829$, while for the running vacuum model
we use $\sigma_{8}=0.750$ (see Refs. \cite{GoSolBas2015,Basrev}). Also, $D(z)$ is
the growth factor normalized to unity at the present time.

As we have already said in the Introduction 
the determination of the
growth index is important in cosmological studies 
because it can be used as a tool toward testing
the validity of general relativity on extragalactic scales.
For those dark energy models that adhere to GR 
and characterized by a constant equation of state parameter it has been found 
that
$\gamma \approx \frac{3(w-1)}{6w-5}$
\cite{Silv94,Wang98,Linder2007,Nes08}, while for the $\Lambda(H)+$GR case 
we obtained $\gamma_{\Lambda_{H}} \approx \frac{6+3\nu}{11-12\nu}$.
Obviously, the growth index reduces to 6/11 for 
the usual $\Lambda$CDM model ($w=-1$ or $\nu=0$).

In the case of extended theories of gravity 
the situation is as follows. Recently, 
for the holographic dark energy models it has been found that 
the asymptotic value of the
growth index is $\gamma \approx 4/7$ \cite{Mehra2015}. 
For the braneworld gravity of Ref.\cite{DGP} we have $\gamma \approx 11/16$
(see also Refs.\cite{Linder2007,Wei08,Gong10,Fu09}), 
for some $f(R)$ gravity models it has been found that
$\gamma \simeq 0.415-0.21z$ (see \cite{Gann09,Tsu09}) and
lastly for the Finsler-Randers cosmological model
Basilakos \& Stavrinos \cite{Basilakos2013}
have shown that $\gamma \approx 9/14$.
Based on the aforementioned results if the derived 
value of $\gamma$ (based on the next generation 
of surveys, cf. {\it Euclid}) shows scale or time dependence or it is 
inconsistent with 6/11 then 
this will be a hint that the nature of DE reflects 
on the physics of gravity.

To conclude, in this work we have analytically 
studied the growth index of matter 
perturbations for the FLRW flat cosmological 
models in which the vacuum energy density is a 
function of the Hubble parameter, namely
$\Lambda(H)=\Lambda_{0}+3\nu(H^{2}-H^{2}_{0})$.
In previous comprehensive studies \cite{BPS09,Grande11,GoSolBas2015}
we have utilized such a dynamical vacuum model in order to investigate
the background expansion and have carefully compared the differences with respect to the concordance $\Lambda$CDM cosmological model.
We believe that the combination of the works of
Refs.\cite{BPS09,Grande11,GoSolBas2015} together with the current article provide a rather
complete investigation of the observational status of the RG running vacuum model both at the background and perturbation levels.

Within this framework, we have calculated for 
the first time (to the best of our knowledge) the
asymptotic value of the growth index,
which is given by $\gamma_{\Lambda_{H}} \approx \frac{6+3\nu}{11-12\nu}$.
Obviously, the obtained formula analytically extends in a very clear way that of the $\Lambda$CDM model, $\gamma_{\Lambda}\approx 6/11$.
In our study we have applied
the two most popular parametrizations for the evolution of the
growth index:
$\gamma(z)=\gamma_{0}+\gamma_{1}y(z)$,
with $y(z)={\rm ln }{\tilde \Omega}_{m}(z)$ and
$y(z)=z/(1+z)$, we have solved the problem analytically 
and we have thus provided
for the first time the coefficients
$\gamma_{0}$ and $\gamma_{1}$ in terms of $\Omega_{m0}$ and $\nu$.
The comparison shows that the above $\gamma(z)$ parametrizations
are practically equivalent and the corresponding evolution of the
growth rate of clustering matches quite well the recent growth data. Finally, we have estimated the numerical corrections that the running $\Lambda(H)$ model could produce on the growth index as compared to the concordance model and pointed out that they could reach, in some cases, the level of a few percent,  hopefully accessible in the future.


\vspace {0.2cm}

\section*{Acknowledgments}
SB also acknowledges support by the Research Center for Astronomy of
the Academy of Athens in the context of the program  ``{\it Tracing
the Cosmic Acceleration}''.
JS has been supported in part by FPA2013-46570 (MICINN),
CSD2007-00042 (CPAN), 2014-SGR-104 (Generalitat de Catalunya) and MDM-2014-0369 (ICCUB). SB is also grateful to the
Department ECM (Universitat de Barcelona) for the hospitality and
support when this work was being started.



\end{document}